\documentclass[prb,twocolumn,superscriptaddress,showpacs,amsmath,amssymb]{revtex4}
\usepackage{graphicx}
\usepackage{dcolumn}
\usepackage{bm}
\usepackage{color}
\usepackage{siunitx}

\begin{document}

\title{From Fe$_3$O$_4$/NiO bilayers to NiFe$_2$O$_4$-like thin films through Ni interdiffusion}

\author{O.~Kuschel}
\affiliation{Department of Physics and Center of Physics and Chemistry of New Materials, Osnabr{\"u}ck University, 49076 Osnabr{\"u}ck, Germany}

\author{R.~Bu\ss}
\affiliation{Department of Physics and Center of Physics and Chemistry of New Materials, Osnabr{\"u}ck University, 49076 Osnabr{\"u}ck, Germany}

\author{W.~Spiess}
\affiliation{Department of Physics and Center of Physics and Chemistry of New Materials, Osnabr{\"u}ck University, 49076 Osnabr{\"u}ck, Germany}

\author{T.~Schemme}
\affiliation{Department of Physics and Center of Physics and Chemistry of New Materials, Osnabr{\"u}ck University, 49076 Osnabr{\"u}ck, Germany}

\author{J.~W\"ollermann}
\affiliation{Department of Physics and Center of Physics and Chemistry of New Materials, Osnabr{\"u}ck University, 49076 Osnabr{\"u}ck, Germany}

\author{K.~Balinski}
\affiliation{Department of Physics and Center of Physics and Chemistry of New Materials, Osnabr{\"u}ck University, 49076 Osnabr{\"u}ck, Germany}

\author{T.~Kuschel}
\affiliation{Center for Spinelectronic Materials and Devices, Department of Physics, Bielefeld University,  Universit\"atsstra\ss e 25, 33615 Bielefeld, Germany}

\author{A.T.~N'Diaye}
\affiliation{Advanced Light Source, Lawrence Berkeley National Laboratory, California 94720, USA}

\author{J.~Wollschl\"ager}\email{jwollsch@uos.de}
\affiliation{Department of Physics and Center of Physics and Chemistry of New Materials, Osnabr{\"u}ck University, 49076 Osnabr{\"u}ck, Germany}

\author{K.~Kuepper}\email{kkuepper@uos.de}
\affiliation{Department of Physics and Center of Physics and Chemistry of New Materials, Osnabr{\"u}ck University, 49076 Osnabr{\"u}ck, Germany}


\date{\today}


\begin{abstract}

Ferrites with (inverse) spinel structure display a large variety of electronic and magnetic properties making some of them interesting for potential applications in spintronics. We investigate the thermally induced interdiffusion of Ni$^{2+}$ ions out of NiO into Fe$_3$O$_4$ ultrathin films resulting in off-stoichiometric nickelferrite-like thin layers. We synthesized epitaxial Fe$_3$O$_4$/NiO bilayers on Nb-doped SrTiO$_3$(001) substrates by means of reactive molecular beam epitaxy. Subsequently, we performed an annealing cycle comprising three steps at temperatures of 400\,$^{\circ}$C, 600\,$^{\circ}$C, and 800\,$^{\circ}$C under an oxygen background atmosphere. We studied the changes of the chemical and electronic properties as result of each annealing step with help of hard x-ray photoelectron spectroscopy and found a rather homogenous distribution of Ni and Fe cations throughout the entire film after the overall annealing cycle. For one sample we observed a cationic distribution close to that of the spinel ferrite NiFe$_2$O$_4$. Further evidence comes from low energy electron diffraction patterns indicating a spinel type structure at the surface after annealing. Site and element specific hysteresis loops performed by x-ray magnetic circular dichroism uncovered the antiferrimagnetic alignment between the octahedral coordinated Ni$^{2+}$ and Fe$^{3+}$ ions and the Fe$^{3+}$ in tetrahedral coordination. We find a quite low coercive field of 0.02\,T, indicating a rather low defect concentration within the thin ferrite films.

\end{abstract}

\maketitle

\section{Introduction}
Iron oxides are of special interest due to a number of astonishing properties in dependence of the Fe valence state and the underlying crystallographic and electronic structure. Magnetite (Fe$_3$O$_4$) is among the most studied ferrites due to its ferrimagnetic ordered ground state with a saturation moment of 4.07$\mu_B$ per formula unit and a high Curie temperature of 860~K for bulk material.\cite{wei29,sli80} This magnetic ground state is accompanied by half metallicity, i.e.\ only one spin orientation is present at the Fermi energy,\cite{kat08} making this material a potential candidate for future spintronic devices with 100\% spin polarization.\cite{bli14,byr15} Magnetite crystallizes in the cubic inverse spinel structure (equal distribution of Fe$^{3+}$ on A and B sites and Fe$^{2+}$ exclusively on B sites) with lattice constant $a$~=~0.8396~nm (space group Fd3m). The oxygen anions form an fcc anion sublattice.\par

Often, Fe$_3$O$_4$ thin films are grown on cubic MgO(001) substrates by various deposition techniques\cite{bal04,ste07,ber12,ber13,moy15b,sch15b}, since the lattice mismatch between the Fe$_3$O$_4$ and MgO(001) ($a$~=~0.42117~nm) is only 0.3\%, comparing the oxygen sublattices. A severe limit of epitaxial thin film growth on MgO substrates is Mg$^{2+}$ segregation into the Fe$_3$O$_4$ film if the substrate temperature is above 250$^{\circ}$C.\cite{kim09} Mg rich interfaces\cite{sha00} and Mg interdiffusion have been studied in detail,\cite{kim08} having significant influence on interface roughness or anti phase boundaries. Thus, the underlying electronic and magnetic structure determining the properties of the magnetite thin film in question or the tunnel magneto resistance in magnetic tunnel junctions with magnetite electrodes.\cite{and97,kal03,wu12,mar15}\par

Potential approach to minimize or suppress Mg segregation, besides rather low substrate temperatures during magnetite growth, is an additional buffer layer, e.g.\ metallic iron\cite{sch15} or NiO\cite{gat05} between the Fe$_3$O$_4$ and the substrate. This approach is also of interest with respect to the possibility for building a full oxidic spin valve making use of the exchange bias between the ferrimagnetic magnetite and the antiferromagnetic nickel oxide.\cite{kel02,gat05,sch15}  The usage of other substrates like SrTiO$_3$ could also prevent Mg interdiffusion. Despite the large lattice mismatch of -7.5\% between the doubled SrTiO$_3$ bulk lattice constant (0.3905~nm) and magnetite it is possible to grow epitaxial Fe$_3$O$_4$ films on the SrTiO$_3$(001) surface.\cite{mon13,rub15} In particular, concerning coupled Fe$_3$O$_4$/NiO bilayers grown on SrTiO$_3$, so far only Pilard \emph{et al.\ } have reported on the magnetic properties of the Fe$_3$O$_4$/NiO interface.\cite{pil07} On the other hand, NiFe$_2$O$_4$ thin films are of huge interest nowadays, since they act as magnetic insulators, which can be used to thermally induce spin currents via the spin Seebeck effect.\cite{mei13,mei15} Furthermore, electrical charge transport and spin currents can be manipulated by the spin Hall magnetoresistance using NiFe$_2$O$_4$ thin films adjacent to nonmagnetic material.\cite{alt13} \par

Here we go beyond describing a model system of two distinct layers with an epitaxial Fe$_3$O$_4$/NiO interface and study the potential Ni$^{2+}$ interdiffusion from a NiO buffer layer into a Fe$_3$O$_4$ top layer as well as NiO surface segregation through the Fe$_3$O$_4$ layer if both NiO buffer layer and Fe$_3$O$_4$ top layer are grown on Nb-doped SrTiO$_3$(001). Knowledge about the modification of the underlying crystallographic, electronic and magnetic structure by Ni interdiffusion is indispensable for potential applications. We also want to learn fundamental aspects especially of Ni$^{2+}$ segregation into epitaxial Fe$_3$O$_4$ thin films, since knowledge of diffusion processes in oxides appear to be still quite rudimentary for many systems.

We perform a systematic three step annealing cycle of Fe$_3$O$_4$/NiO bilayers after synthesis and simultaneously investigating surface crystallographic and "bulk" electronic structure changes by means of low energy electron diffraction (LEED) and hard x-ray photoelectron spectroscopy (HAXPES). Furthermore, we carry out structural analysis before and after the overall annealing cycle employing x-ray reflectivity (XRR) and synchrotron based x-ray diffraction (SR-XRD), as well as element and site specific x-ray magnetic circular dichroism (XMCD) after the overall annealing cycle to analyze the resulting magnetic properties in detail.

\section{Experimental details}
Two samples with \(\mathrm{Fe_3O_4}\)/\(\mathrm{NiO}\) ultra thin film bilayers on conductive \SI{0.05}{{wt.}~\percent} \(\mathrm{Nb}\)-doped \(\mathrm{SrTiO_3}\)(001) substrates have been prepared, using the technique of reactive molecular beam epitaxy (RMBE).
The substrates have been supplied with a polished surface and were annealed at \SI{400}{\celsius} for one hour in an oxygen atmosphere of \SI{1d-4}{\milli\bar} prior to deposition.
During film growth, the oxygen pressure was kept at \SI{1d-5}{\milli\bar} for \(\mathrm{NiO}\) and \SI{5d-6}{\milli\bar} for \(\mathrm{Fe_3O_4}\), while the substrate was heated to \SI{250}{\celsius}.
One sample has been created with a 5.6\,nm \(\mathrm{NiO}\) film (sample A) and the other with a 1.5\,nm \(\mathrm{NiO}\) film (sample B). Thereafter, 5.5\,nm thick \(\mathrm{Fe_3O_4}\) films were deposited on the NiO films.
Substrate preparation, film stoichiometry and surface structure have been monitored \textit{in-situ} by x-ray photoelectron spectroscopy (XPS) using Al K$\alpha$ radiation and LEED, respectively.

The samples were transported under ambient conditions to the Diamond Light Source (DLS) synchrotron, where the effects of annealing on the bilayer system were studied at beamline I09 by heating the samples in three steps at \SI{400}{\celsius}, \SI{600}{\celsius}, and \SI{800}{\celsius} for 20 to 30 minutes in an oxygen atmosphere of \SI{5d-6}{\milli\bar}. Prior to and after the annealing studies XRR measurements at 2.5\,keV photon energy were made to determine the film thickness. After each annealing step, the films were studied \textit{in-situ} by soft x-ray photoemission and HAXPES to clarify the chemical composition in the surface near region and in the bulk region, respectively. In addition, LEED measurements were performed to check the crystallinity of the individual layers of the NiO/Fe$_3$O$_4$ bilayer.

For HAXPES an energy of \(h\nu = \SI{5934}{\electronvolt}\) was used, creating photoelectrons with high kinetic energy, which allows a higher probing depth compared to soft x-ray photoemission (\(h\nu = \SI{1000}{\electronvolt}\)).
The information depth, from which \SI{95}{\percent} of the photoelectrons originate, is defined as
\begin{equation}
ID(95) = -\lambda \cos \varphi \ln(1-95/100)\text{,}
\end{equation}
with the inelastic mean free path \(\lambda\) and the off-normal emission angle \(\varphi\).\cite{id}
The maximum information depth for the \(\mathrm{Fe~2p}\) core level for HAXPES and soft x-ray photoemission measurements is \(\SI{22}{\nano\meter}\) and \(\SI{2.5}{\nano\meter}\), respectively, estimating \(\lambda\) by the TPP-2M formula.\cite{imfpVII}
As the beamline features a 2D photoelectron detector, which can be operated in an angular mode, photoelectron spectra at different emission angles were acquired, each with an acceptance angle of \SI{\sim 7}{\degree}.

Subsequently, structural analysis of the annealed films was performed using SR-XRD, while the resulting film thickness and layer structure of this films were determined by means of lab based XRR using Cu K$_\alpha$ radiation. SR-XRD experiments have been carried out \textsl{ex-situ} at \mbox{PETRA III} beamline P08 (DESY, Germany) using a photon energy of 15\,keV. In both cases the measurements were performed in $\theta\,-\,2\theta$ diffraction geometry.
For the analysis of all XRR experiments an in-house developed fitting tool based on the Parratt algorithm \cite{parratt} and N\'{e}vot-Croce roughness model \cite{nevot} was used. The SR-XRD measurements were analyzed by calculating the intensity distribution within the full kinematic diffraction theory to fit the experimental diffraction data.

XMCD spectroscopy was performed at the Fe L$_{2,3}$ and Ni L$_{2,3}$ edges with the samples at room temperature at beamline 6.3.1 of the Advanced Light Source, Lawrence Berkeley Laboratory. We have utilized total electron yield (TEY) as detection mode. The external magnetic field of 1.5 T has been aligned parallel to the x-ray beam and been switched for each energy. The angle between sample surface and x-ray beam has been chosen 30$^{\circ}$. The resolving power of the beamline has been set to $E/ \Delta E$ $\sim$2000, the degree of circular polarization has been about 55\%. For the analysis of the Fe L$_{2,3}$ XMCD spectra, we have performed corresponding model calculations within the atomic multiplet and crystal field theory including charge transfer using the program \textsc{CTM4XAS}.\cite{dgr05,sta10}

\section{Results}

\subsection{Surface characterisation}
\begin{figure}[ht]
\centering
	\includegraphics[width=0.85\linewidth]{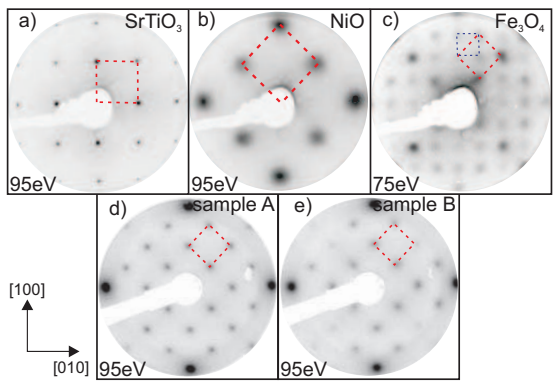}
	\caption{LEED patterns taken after a) preparation of \(\mathrm{SrTiO_3}\) substrate, b) deposition of \(\mathrm{NiO}\), c) deposition of \(\mathrm{Fe_3O_4}\), d) and e) after annealing at \SI{800}{\celsius} of sample A and sample B, respectively . Marked with red squares are the respective \((1\times 1)\) surface unit cells in reciprocal space. The blue square indicates the (\(\sqrt{2}\times\sqrt{2}\))R45$^\circ$ superstructure typical for magnetite.}
\label{fig:LEED}
\end{figure}

After cleaning of the \(\mathrm{SrTiO_3}\) substrates the XPS shows chemically clean substrates without carbon contamination (not shown here). The LEED pattern shows very sharp diffraction spots of a (\(1\times 1\)) surface with square structure and negligible background intensity (cf. Fig.~\ref{fig:LEED}a), indicating a clean (001) oriented surface with long range structural order.


The LEED image recorded after RMBE of \(\mathrm{NiO}\) also exhibits a (\(1\times 1\)) structure. The pattern, however (cf. Fig.~\ref{fig:LEED}b), is rotated by \SI{45}{\degree} and $\sim \sqrt{2}$ times larger than the pattern of the SrTiO$_3$(001) substrate as expected for the \(\mathrm{NiO}\)(001) surface unit cell.
The broadening of the diffraction spots is most likely caused by defects due to relaxation processes induced by the high lattice misfit of \SI{-6.9}{\percent} for \(\mathrm{NiO}\)(001) compared to \(\mathrm{SrTiO_3}\)(001).

The LEED pattern of the as prepared \(\mathrm{Fe_3O_4}\) film (cf. Fig.~\ref{fig:LEED}c) reveals a (\(1\times 1\)) surface structure with doubled periodicity compared to \(\mathrm{NiO}\), as the real space lattice constant of the magnetite inverse spinel structure is about twice as large, giving a lattice misfit of only \SI{0.7}{\percent} for \(\mathrm{Fe_3O_4}\)(001) on \(\mathrm{NiO}\)(001).
Furthermore, additional diffraction spots of a (\(\sqrt{2}\times\sqrt{2}\))R45$^\circ$ superstructure can be seen, which is unique for well-ordered \(\mathrm{Fe_3O_4}\)(001) surfaces.\cite{feosurfacetermination, korecki, pentcheva}

These results indicate a cube-on-cube growth for both, \(\mathrm{NiO}\) and \(\mathrm{Fe_3O_4}\) films. Additionally, the Ni 2p and Fe 2p XPS spectra taken directly after preparation of each film (not shown here) exhibit a characteristic shape for a Ni$^{2+}$ and a mixed Fe$^{2+}$/Fe$^{3+}$ valence state, respectively. Thus, combining the results from XPS and LEED, we can conclude that the as-prepared films are consisting of stoichiometric Fe$_3$O$_4$/NiO bilayers.

The first annealing step at \SI{400}{\celsius} only removed surface contaminations from the transport, without effecting the initial layer structure of the sample. Soft x-ray photoemission measurements show a characteristic Fe~2p signal indicating a \(\mathrm{Fe_3O_4}\) stoichiometry. Furthermore, no Ni 2p signal was visible due to the small information depth demonstrating that neither Ni diffused into the Fe$_3$O$_4$ film nor that the Fe$_3$O$_4$ film was deconstructed.

After the annealing step at \SI{600}{\celsius} and \SI{800}{\celsius} a distinctive satellite typical for trivalent iron becomes visible between the Fe~2p$_{1/2}$ and Fe~2p$_{3/2}$ peaks for soft x-ray photoemission measurements. This indicates a deficiency of divalent iron in the magnetite layer. Furthermore, the spectra show an intense Ni~2p signal pointing to a possible deconstruction or a formation of nickel ferrite as a result of intermixing.

LEED patterns (cf. Fig.~\ref{fig:LEED}d, e) taken after the entire annealing experiments show a clear (\(1\times 1\)) surface structure for both samples. Sample A, however, exhibits sharper reflexes than sample B. This structure corresponds to the inverse spinel surface structure described above for magnetite, but without the (\(\sqrt{2}\times\sqrt{2}\))R45$^\circ$ superstructure of the \(\mathrm{Fe_3O_4}\)(001) surface. Therefore, it also can be attributed either to a defect rich magnetite surface, or the formation of several iron oxide species but also to a NiFe$_2$O$_4$ surface.

\subsection{XRR}

\begin{figure}[ht]
\centering
	\includegraphics[width=0.85\linewidth]{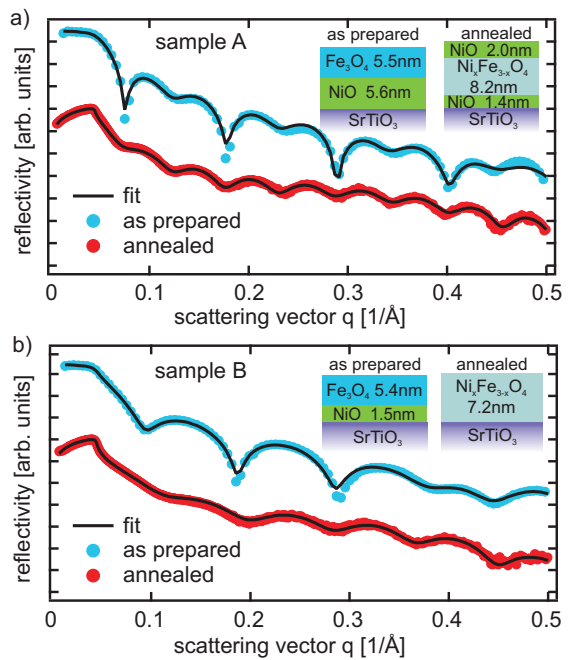}
	\caption{Reflectivity curves and calculations from XRR measurements before and after the annealing experiments a) for sample A and b) for sample B. The insets show the underlying models.}
\label{fig:XRR}
\end{figure}

Fig.~\ref{fig:XRR} shows the measured and calculated XRR intensities obtained at DLS prior to the the annealing experiments for both samples. The XRR intensity obtained from sample A clearly shows the beating of two layers with almost identical thickness while the intensity obtained from sample B points to two layers with very different thickness. For the calculation of the intensity distributions and the exact layer structure a basic model was used (inset Fig.~\ref{fig:XRR}), consisting of a magnetite film on top of a NiO layer on a SrTiO$_3$ substrate. For both samples the data show well defined intensity oscillations pointing to a double layer structure and flat homogeneous interfaces and films.


The measured and calculated XRR intensities of the annealed samples as well as the used model are also presented in Fig.~\ref{fig:XRR}. For both samples the XRR shows clear intensity oscillations with a changed periodicity compared to the as prepared films. Taking into account the electron densities and layer structures obtained from XRR this effect can be attributed to an intermixing of the two initial oxide layers. In case of sample A a three layer model was necessary to describe the data after annealing (cf.~Fig.~\ref{fig:XRR}a). The first layer on top of the substrate is a thin nickel oxide layer, the second layer is a 8.2~nm thick nickel ferrite film and the third layer on top of the nickel ferrite film consists again of nickel oxide.\par

The model parameters of the upper NiO layer indicate a deconstructed film or island formation on the surface. However, sample B is modeled with only a single 7.2~nm thick nickel ferrite film on top of the substrate (cf.~Fig.~\ref{fig:XRR}b). For both samples the thicknesses of the residual films coincide almost with the sum of the initial thicknesses of the \(\mathrm{Fe_3O_4}\) and NiO films. The slightly increase of the thickness can be attributed to a volume increase of $\sim 8\%$ due to the formation of nickel ferrite.


\subsection{HAXPES}


\begin{figure}[ht]
\centering
 \includegraphics[width=0.8\linewidth]{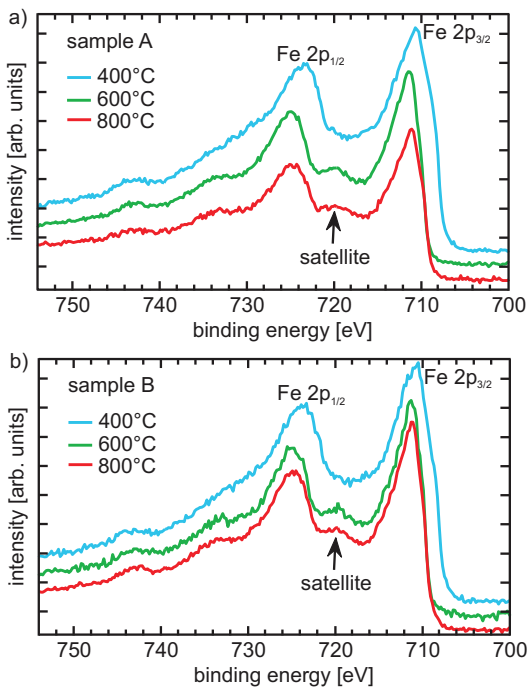}
\caption{HAXPES spectra of Fe~2p core level at \SI{10}{\degree} off-normal photoelectron emission after annealing at different temperatures a) for sample A and b) for sample B.}
\label{fig:Fe2p}
\end{figure}


\begin{figure}[ht]
\centering
 \includegraphics[width=0.8\linewidth]{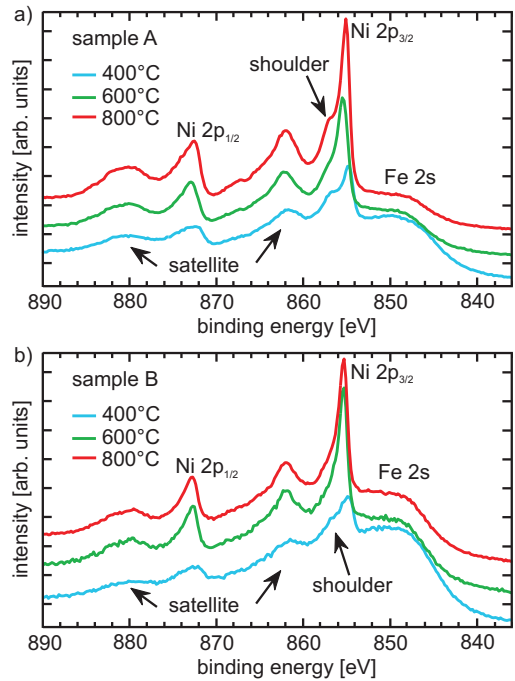}
\caption{HAXPES spectra of Ni~2p core level at \SI{10}{\degree} off-normal photoelectron emission after annealing at different temperatures a) for sample A and b) for sample B.}
\label{fig:Ni2p}
\end{figure}

In contrast to soft x-ray photoemission, HAXPES measurements allow to identify the valence states and chemical properties not only at the surface near region but with bulk sensitivity due to higher excitation energy and, thus, increased
information depth.

Fig.~\ref{fig:Fe2p} shows the HAXPES spectra for the \(\mathrm{Fe~2p}\) core level, which is split into the \(\mathrm{Fe~2p_{1/2}}\) and \(\mathrm{Fe~2p_{3/2}}\) peaks. Spectra recorded after each annealing step for both samples are presented.
The shape of the spectra is determined by the relative fraction of Fe valence states, which is used to identify the material composition\cite{ironOxideValances1}.
After the initial annealing step at \SI{400}{\celsius}, there is no satellite peak visible between the two main peaks, indicating stoichiometric \(\mathrm{Fe_3O_4}\) for both samples.
After the second and third annealing step, at \SI{600}{\celsius} and \SI{800}{\celsius}, respectively, a satellite peak becomes visible between the two main peaks for both samples.
As it resides on the side of the \(\mathrm{Fe~2p_{1/2}}\) peak, it indicates a deficiency of \(\mathrm{Fe^{2+}}\) ions in favor of \(\mathrm{Fe^{3+}}\) ions compared to the initial magnetite stoichiometry.

Fig.~\ref{fig:Ni2p} shows the photoelectron spectra for the \(\mathrm{Ni~2p_{1/2}}\) and \(\mathrm{Ni~2p_{3/2}}\) core level of both samples.
The spectra after the annealing step at \SI{400}{\celsius} show a shoulder on the high binding energy side of the \(\mathrm{Ni~2p_{3/2}}\) peak, which is typical for \(\mathrm{NiO}\).\cite{nickelPeakshape}
This shoulder almost completely disappears after annealing at \SI{600}{\celsius} of both samples.
\citeauthor{NFOpeakshape}\cite{NFOpeakshape} identified such a peak shape without a satellite for the spinel type material \(\mathrm{NiFe_2O_4}\).
Therefore, an exchange of \(\mathrm{Fe^{2+}}\) ions with \(\mathrm{Ni^{2+}}\) ions in the \(\mathrm{Fe_3O_4}\) spinel structure through interdiffusion seems to be likely.\cite{theironoxides}
For sample B, the peak shape does not change with the next annealing step at \SI{800}{\celsius} (cf.~Fig.~\ref{fig:Ni2p}b). However, for sample A the shoulder on the high binding energy side observed for the initial bilayer system re-appears (cf.~Fig.~\ref{fig:Ni2p}a), suggesting the formation of \(\mathrm{NiO}\) like structures, which is consistent with the NiO formation at the surface seen in the XRR measurements.

\begin{figure}[ht]
\centering
    \includegraphics[width=\linewidth]{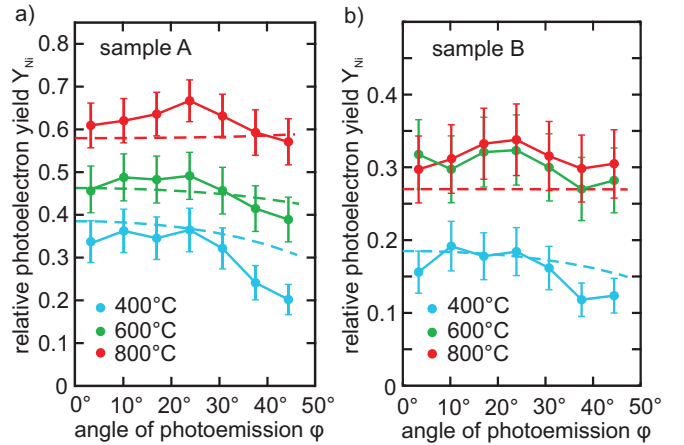}
\caption{Relative photoelectron yield at different off-normal emission angles a) for sample A and b) for sample B. The dashed lines show the calculated intensities using the models obtained from XRR analysis.}
\label{fig:HAXPESratio}
\end{figure}

Additionally, a quantitative analysis of the photoelectron spectra was performed to prove the possible formation of nickel ferrite.
After subtracting a Shirley background, the intensities \(I_{\mathrm{Fe}}\) and \(I_{\mathrm{Ni}}\) of the \(\mathrm{Fe~2p}\) peaks and the \(\mathrm{Ni~2p_{1/2}}\) peak (due to the overlap with the \(\mathrm{Fe~2s}\), the \(\mathrm{Ni~2p_{3/2}}\) peak has not been considered) have been numerically integrated.
From these results, the relative photoelectron yield
\begin{equation}
Y_{\mathrm{Ni}} = \frac{I_{\mathrm{Ni}}/\sigma_{\mathrm{Ni}}}{I_{\mathrm{Ni}}/\sigma_{\mathrm{Ni}} + I_{\mathrm{Fe}}/\sigma_{\mathrm{Fe}}} = \frac{N_{\mathrm{Ni}}}{N_{\mathrm{Ni}} + N_{\mathrm{Fe}}\cdot C(\varphi)}
\label{yield_eq}
\end{equation}
of Ni has been calculated, using the differential photoionization cross sections \(\sigma\) reported by \citeauthor{pad} \cite{pad}
\citeauthor{newberg} derived, that this yield is equal to the atomic ratios\cite{newberg}, but with a factor \(C(\varphi)\), that depends on the angle of photoemission.
The resulting yields from different detection angles are plotted in Fig.~\ref{fig:HAXPESratio}.
The curves from the data of the first annealing steps show for both samples a decreasing yield for higher emission angles as indicated by the blue dashed lines. This behavior points to an intact stack of oxide films due to a longer pathway of the photoelectrons for higher emission angles. The lines are calculated for a stack of two separated \(\mathrm{Fe_3O_4}\)/\(\mathrm{NiO}\) films using the thicknesses obtained from XRR analysis.
With the successive annealing steps, the photoelectron yield from Ni increases, which indicates that there is diffusion of Ni into the \(\mathrm{Fe_3O_4}\) film and/or Fe into the \(\mathrm{NiO}\) film.

 In case of sample A the intensity ratios (Fig.~\ref{fig:HAXPESratio}a) show a continuous increase of the nickel intensity with higher annealing temperature. The calculation of the photoelectron yield after the third annealing step at 800~$^\circ$C (dashed red line) was done using the layer structure and thicknesses obtained from the XRR analysis (cf.~inset Fig.~\ref{fig:XRR}a). This model is based on a stochiometric 8.2~nm thick NiFe$_{2}$O$_4$ film between two NiO films.
Since there is no evidence of NiO in the Ni~2p HAXPES spectra after annealing at 600~$^\circ$C, a model consisting of a stoichiometric 8.2~nm thick NiFe$_{3}$O$_4$ on top of a 3.4~nm thick NiO layer was used (green dashed line Fig.~\ref{fig:HAXPESratio}a). With further annealing at 800~$^\circ$C more Ni atoms are diffusing/transported through the nickel ferrite layer to the very surface forming NiO as detected by XRR and HAXPES (see discussion of Ni 2p$_{3/2}$ shoulder) after the entire annealing experiment. This segregation behavior of Ni and the formation of NiO at the surface is explained by its lower surface energy of 0.863~J/m$^2$ compared to the surface energy of 1.235~J/m$^2$ for NiFe$_2$O$_4$(001).\cite{brien}

In case of sample B one can conclude that a single homogeneous film was formed by the interdiffusion process already after the second annealing step. Its stoichiometry does not change from the second to the third annealing step (cf. Fig.~\ref{fig:HAXPESratio}b). The ratio of Ni and Fe, assuming a complete intermixing, can be determined from equation~(\ref{yield_eq}), as then the angular factor \(C(\varphi)\equiv1\). The amount of nickel and iron does not match the ratio of 1~:~2 for stoichiometric nickel ferrite, but is 1~:~2.6 for the sample B indicating an excess of Fe atoms. The experimental data are in good agreement with the calculated behavior (dashed red line) for a homogeneously mixed single layer. Thus, the resulting stoichiometry of the sample B is Ni$_x$Fe$_{3-x}$O$_4$ with $x\,=\,0.83$.

All calculations of the HAXPES intensity indicate the same layer structure and thicknesses as obtained from the XRR measurements were used, indicating a consistent model.


\subsection{SR-XRD}

\begin{figure}[ht]
\centering
    \includegraphics[width=0.85\linewidth]{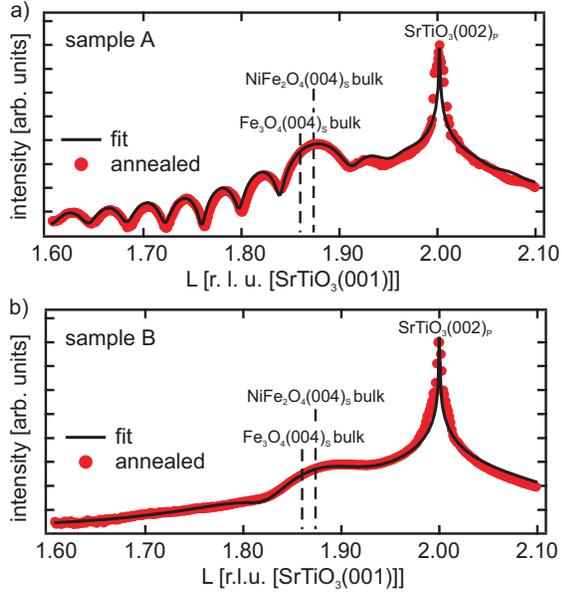}
\caption{SR-XRD measurements along (00\textit{L}) direction and calculated intensities. For the calculation the same model as obtained from the analysis of the XRR was used (cf.~inset Fig.~\ref{fig:XRR}).}
\label{XRD}
\end{figure}

Fig.~\ref{XRD} shows SR-XRD measurements and calculated intensity of the crystal truncation rod (CTR) along (00\textsl{L}) direction close to the SrTiO$_3$(002)$_P$ and spinel (004)$_S$ Bragg peak for both samples after annealing. Here, \textsl{L} denotes vertical scattering vector in reciprocal lattice units (r.l.u.) with respect to the layer distance of the SrTiO$_3$(001) substrate. Indices \textit{P} and \textit{S} indicate the bulk notation for perovskite type and spinel type unit cells, respectively. The structural parameters, e.g. vertical layer distances, are determined by analyzing the CTRs applying full kinematic diffraction theory. For the analysis, the same layer model as for the XRR calculations was used to describe the data (cf. inset Fig.~\ref{fig:XRR}a, b). For both samples a clear peak from the SrTiO$_3$(001) substrate at \textsl{L}\,=\,2 and a broad Bragg peak originating from the oxide film around $\textsl{L} \approx 1.87$ is observed. Additionally, for sample A clear oscillations close to the Bragg peak of the oxide film (Laue fringes) are visible which can be clearly attributed to the nickel ferrite layer indicating a well ordered homogeneous film of high crystalline quality. Furthermore, the vertical lattice constant of sample A obtained from curve fitting is $c$~=~0.8334\,nm and is in good agreement with the bulk value of NiFe$_2$O$_4$ (a$_{bulk}$~=~0.8339\,nm).


For sample B the oscillations completely vanish, pointing to an inhomogeneous film. This effect is possibly caused by the excess of Fe atoms in the film as observed by HAXPES. In addition, the vertical lattice constant c~=~0.8304\,nm  obtained from the calculations confirms the presence of a strongly distorted structure of the annealed film, since it notably comes below the value of bulk NiFe$_2$O$_4$.


\subsection{XMCD}

\begin{figure}[hbt]
    \centering
        \includegraphics[width=7.9cm]{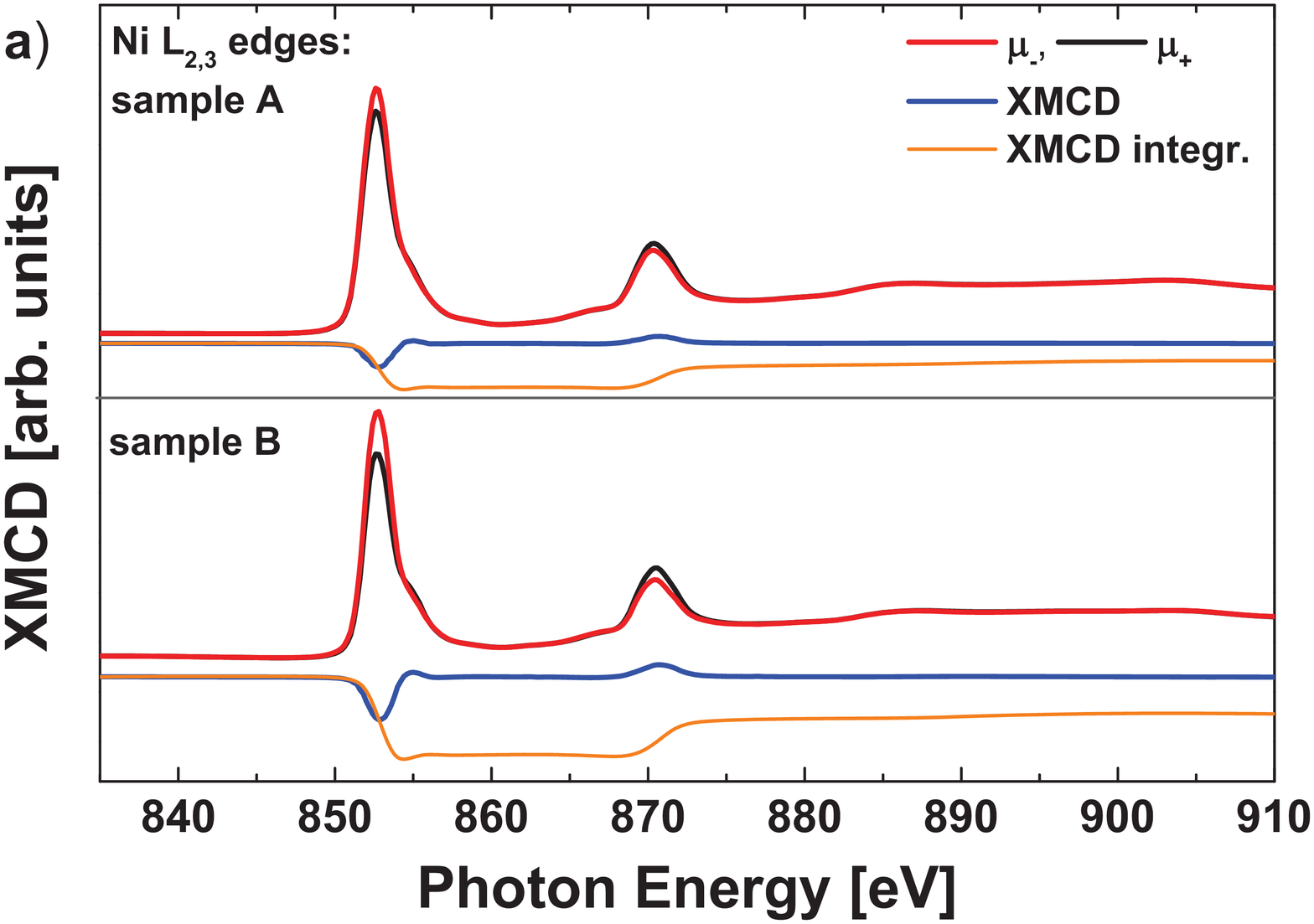}
        \includegraphics[width=7.9cm]{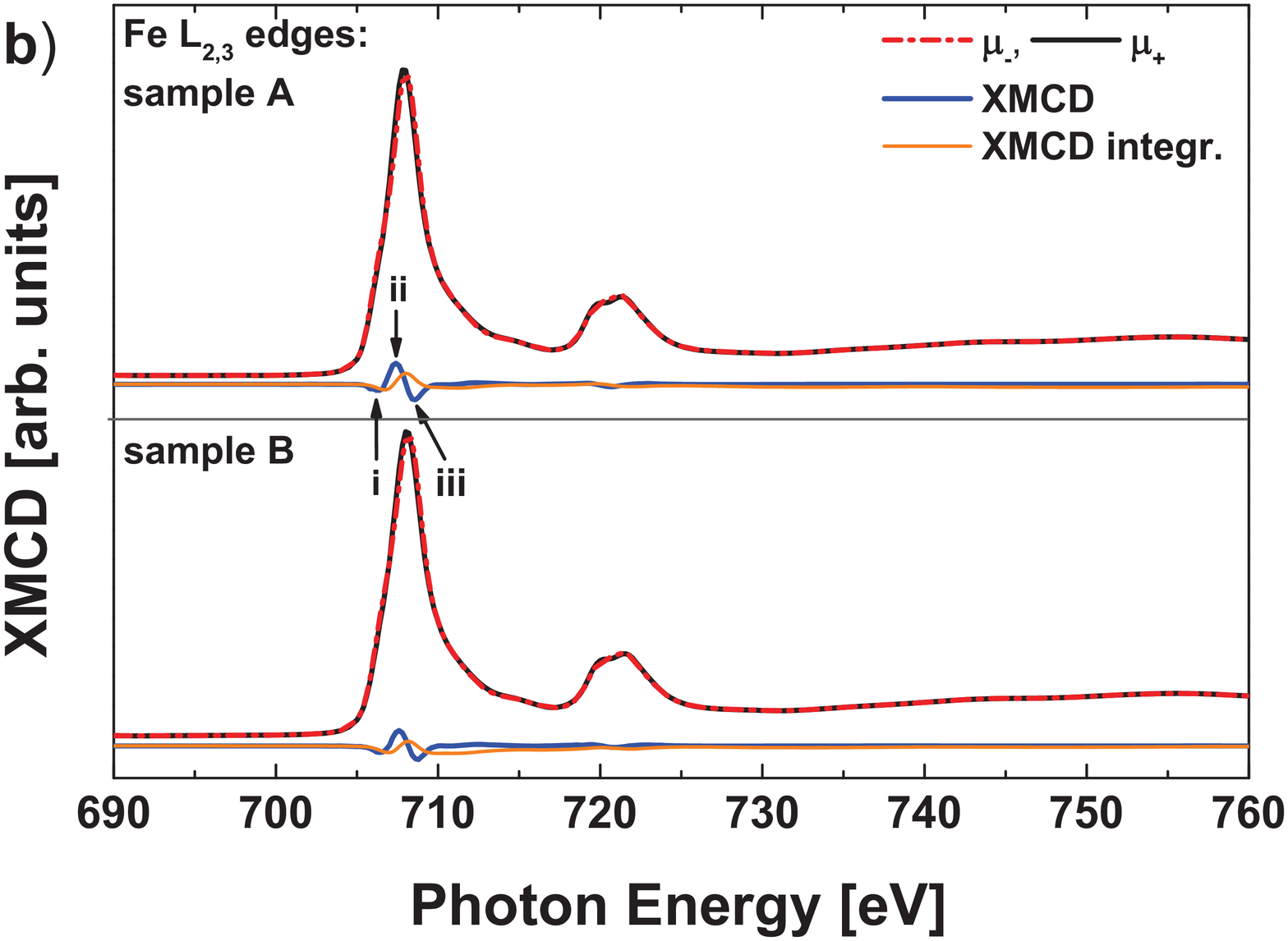}
        \includegraphics[width=7.9cm]{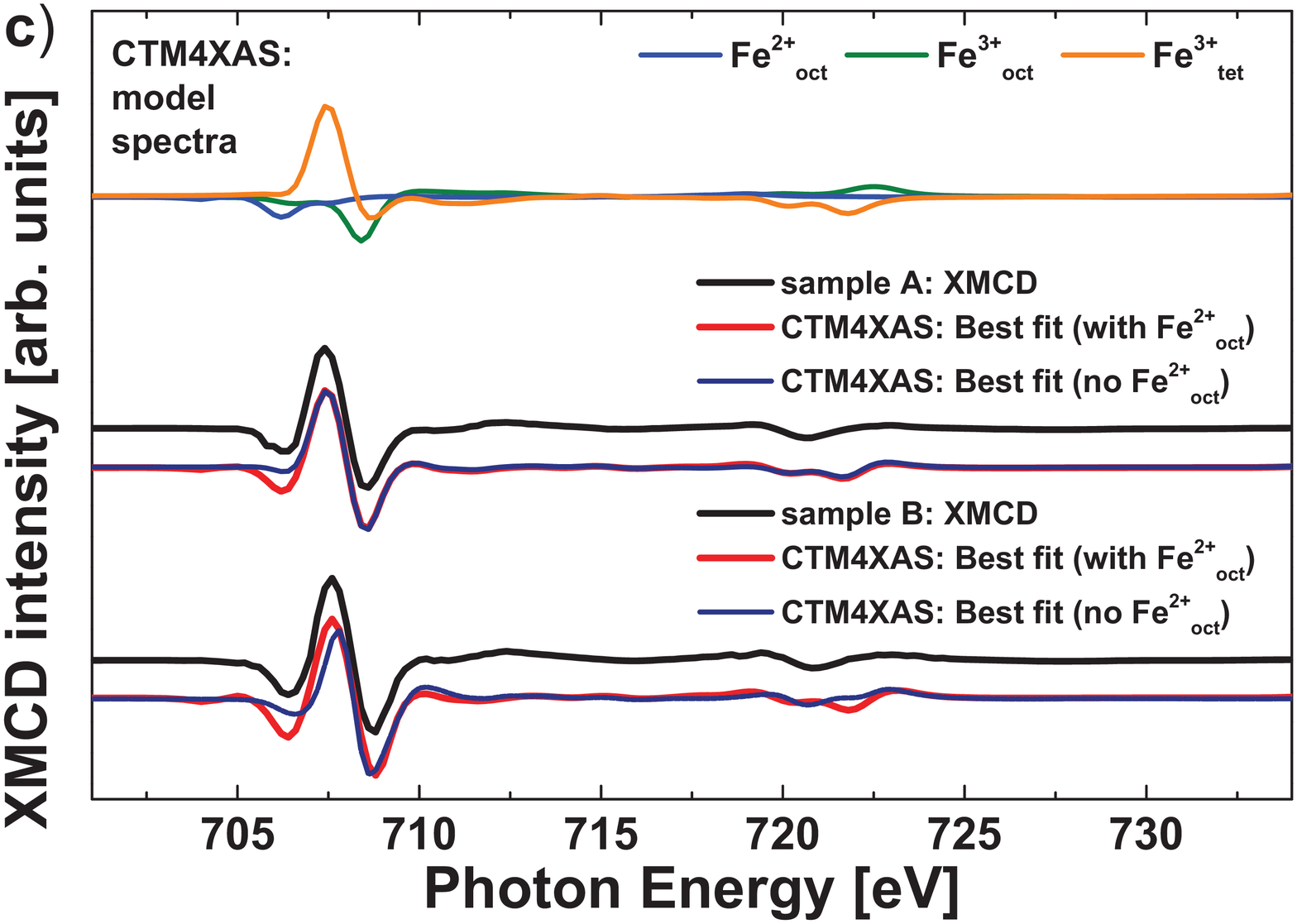}
    \caption{\label{XMCD1} a) Ni L$_{2,3}$-XMCD spectra of samples A and B. b) Fe L$_{2,3}$-XMCD spectra of samples A and B. c) Experimental Fe L$_{2,3}$ edge XMCD of samples A and B and the corresponding XTM4XAS fits with and without consideration of octahedral coordinated Fe$^{2+}$ ions present.}
\end{figure}

XMCD has been employed after the overall annealing cycle to analyze the resulting magnetic properties element specifically after annealing at 800~$^\circ$C. Fig.~\ref{XMCD1} depicts the XMCD spectra of samples A and B performed at the Fe L$_{2,3}$ and Ni L$_{2,3}$ edges, respectively. Both samples show a strong Ni dichroic signal (cf.~Fig.~7a), and in order to extract the spin magnetic moments we use the spin sum rule developed by Chen \emph{et al.}\cite{che95} The number of holes are determined from the charge transfer multiplet simulations for each sample. We also account for the core hole interactions which mix the character of the L$_3$ and L$_2$ edges\cite{ter96,pia09} by considering the spin sum rule correction factors obtained by Teramura \emph{et al.}\cite{ter96} We find a Ni spin moment of 0.51$\mu_B$ per Ni atom and an orbital contribution of 0.053$\mu_B$/Ni atom summing up to a total Ni moment of 0.56 $\mu_B$ for sample A. In case of sample B we derive $m_{spin}$ = 0.91$\mu_B$/Ni atom, $m_{orb}$ = 0.122$\mu_B$/Ni atom, and hence a total Ni moment of 1.03$\mu_B$ per formula unit. The latter value is rather close to that recently found by Klewe \emph{et al.}\cite{kle14} on a stoichiometric NiFe$_2$O$_4$ thin film.\par

Turning to the Fe moments we find strong indications that our heat and diffusion experiments lead to a Ni$_x$Fe$_{3-x}$O$_4$ layer or cluster formation in both samples. Since we obtain $m_{spin}$ = -0.028 (+0.11)$\mu_B$/Fe atom and $m_{orb}$ = +0.015 (+0.007)$\mu_B$/Fe atom at the Fe sites of sample A (sample B) we find very small net contributions to the overall magnetic moments. In comparison Klewe \emph{et al.}\cite{kle14} found an iron spin moment of around 0.1$\mu_B$/Fe atom and a further orbital contribution of around 10-15\% of that value. This indicates an (almost complete) structural inversion of the prior bilayer system, i.e. the iron ions occupy in equal parts octahedral and tetrahedral positions within the crystal. Since the moments of these octahedrally and tetrahedrally coordinated cations are aligned antiparallel the moments cancel each other nearly out in a perfect inverse spinel structure.

Fig.~\ref{XMCD1}c presents the charge transfer multiplet calculations for the single iron cations in octahedral and tetrahedral coordination as well as the best fits to the experimental Fe L$_{2,3}$-XMCD spectra of sample A and B with (red) and without (blue) consideration of Fe$^{2+}_{oct}$ ions. The resulting lattice site occupancies are 16.3\% Fe$^{2+}_{oct}$, 32.2\% Fe$^{3+}_{oct}$, 51.5\% Fe$^{3+}_{tet}$ (42.6\% Fe$^{3+}_{oct}$, 57.4\% Fe$^{3+}_{tet}$) for sample A, and 24.0\% Fe$^{2+}_{oct}$, 31.5\% Fe$^{3+}_{oct}$, 44.5\% Fe$^{3+}_{tet}$ (55.6\% Fe$^{3+}_{oct}$, 44.4\% Fe$^{3+}_{tet}$) for sample B including (not including) Fe$^{2+}_{oct}$ ions into the respective fit. The result that for sample A over 50\% are in Fe$^{3+}_{tet}$ coordination as to the calculations also corresponds with the small negative spin moment determined by the spin sum rule.\par

From the overall multiplet fits (Fig.~\ref{XMCD1}c) one can clearly see that feature $i$ (Fig.~\ref{XMCD1}b) is very small if Fe$^{2+}_{oct}$ cations are not explicitly considered in the respective simulations. The origin of this feature in ferrites with inverse spinel structure other than magnetite is still not entirely understood.\cite{pat02,kle14,hop15} In both Fe L$_{2,3}$-XMCD spectra of samples A and B peak $i$ is significantly smaller than results obtained very recently on NiFe$_2$O$_4$ thin films grown by pulsed laser deposition (PLD),\cite{hop15} but somewhat more intense than it is in the result of Klewe \emph{et al.}\cite{kle14} Also their corresponding multiplet simulation resembles our approach (not considering the Fe$^{2+}_{oct}$ sites) rather well. The presence of peak $i$ in the Fe L$_{2,3}$-XMCD of sample B can at least partly be explained by the lack of Ni$^{2+}_{oct}$ ions as to the HAXPES measurements. Since peak $i$ also occurs in XMCD experiments on bulk material\cite{pat02} one can think about several additional reasons about the presence of some Fe$^{2+}_{oct}$ ions. For instance, a small fraction of the Ni ions might be present in form of Ni$^{3+}$ or coordinated on tetrahedral sites as result of the interdiffusion process. Despite the fact that Ni$^{2+}$ prefers octahedral coordination, even measurements on NiFe$_2$O$_4$ bulk crystals indicate a few of the Ni ions to be on tetrahedral sites.\cite{pat02} Furthermore, oxidation or reduction of a fraction of the Fe at the very surface of the thin films can not be entirely excluded as the probing depth of the total electron yield is only around 2nm at the Fe L$_{2,3}$ and Ni L$_{2,3}$ resonances of oxides.\cite{pat02,gom14}

\begin{figure}[hbt]
    \centering
        \includegraphics[width=7.9cm]{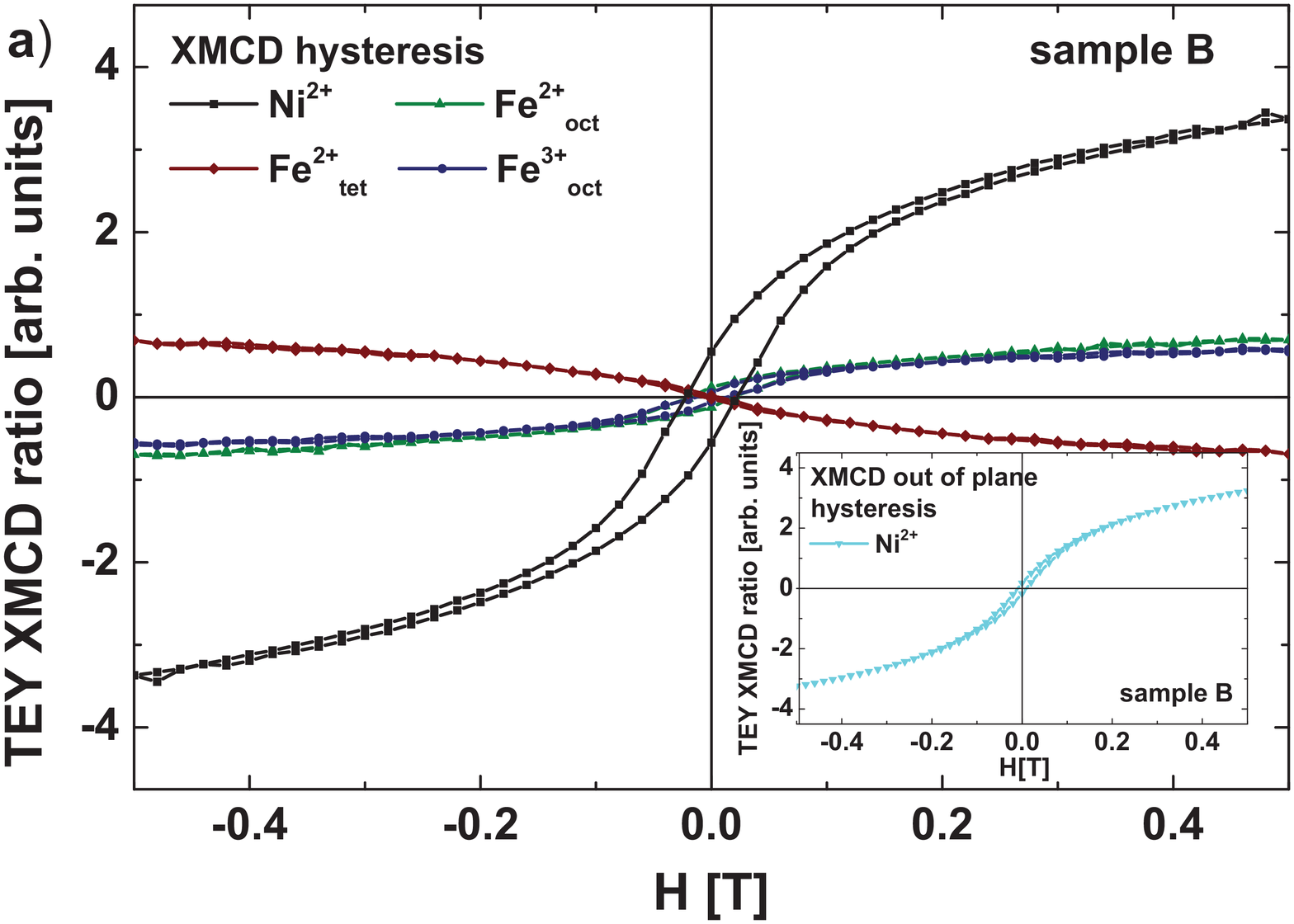}
        \includegraphics[width=7.9cm]{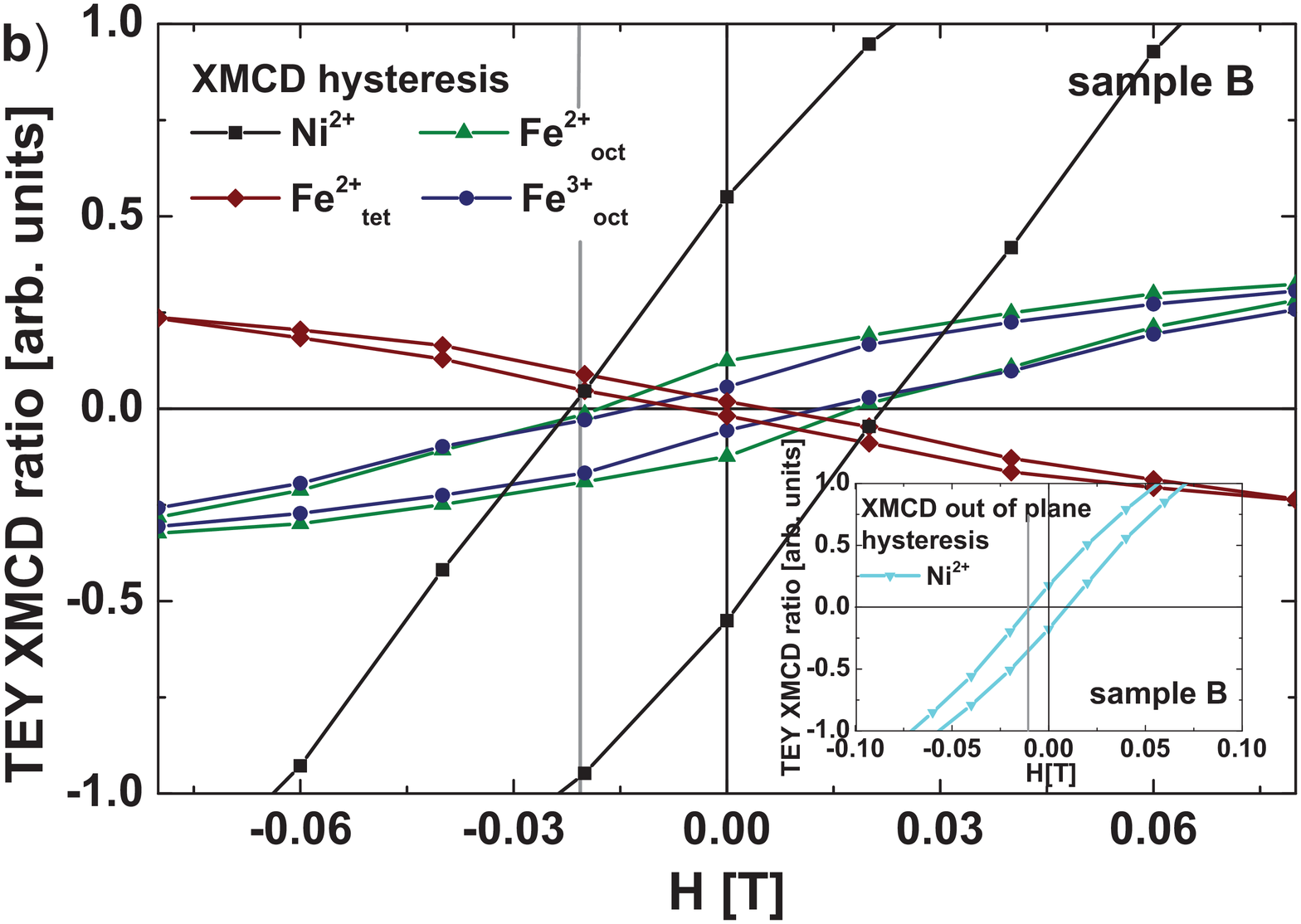}
    \caption{\label{XMCD2} (a) Element and site specific hysteresis loops of the Ni L$_{3}$- and Fe L$_3$ intensities of sample B. (b) Expanded view of the loops near H=0~T. Insets show the Ni hysteresis loop measured in perpendicular (out of plane) geometry.}
\end{figure}

For sample B we also recorded element specific hysteresis loops at the Ni L$_3$ edge and the site specific loops at Fe L$_3$ resonances for peaks $i-iii$ (cf.~Fig.~\ref{XMCD1}b). Fig.~\ref{XMCD2} displays the resulting magnetization loops. One can see the ferrimagnetic ordering between the Fe$^{3+}_{tet}$ cations and the other Fe and Ni cations. For all octahedrally coordinated cations we probe an in-plane coercive field $H_c$ of around 0.02~T, whereas the Fe$^{3+}_{tet}$ cations exhibit a closed, paramagnetic magnetization curve. In out-of plane configuration we only probed the Ni sites and find a $H_c$  of around 0.01~T (see insets in Fig.~\ref{XMCD2}). This is a significant different result compared to recently reported values of $H_c$=0.1~T or more for NiFe$_2$O$_4$ thin films.\cite{jaf12,kle14,hop15} A number of reasons might be responsible for a strongly increased $H_c$ such as exchange coupled grains\cite{jaf12} or a high defect density.\cite{kle14} On the other hand, similar values for the coercive field measured here have been found on polycrystalline as well as epitaxial Ni$_x$Fe$_{3-x}$O$_4$ thin films.\cite{jin10} The bulk value of NiFe$_2$O$_4$ has been reported to be 0.01~T\cite{sha15}  which is close to the values obtained here.


\section{Conclusion}

We investigated the modification of the crystallographic, electronic, and magnetic properties of Fe$_3$O$_4$/NiO-bilayers due to thermally induced interdiffusion of Ni ions out of the NiO layer into the magnetite film. We annealed two bilayers, sample A (B) comprising initially 5.6nm (1.5nm) NiO and 5.5nm (5.4nm) Fe$_3$O$_4$ in three steps \`{a} 20~-~30 minutes in an oxygen atmosphere of \SI{5d-6}{\milli\bar}. LEED demonstrates the extinction of the magnetite specific (\(\sqrt{2}\times\sqrt{2}\))R45 superstructure, however, a spinel like (\(1\times 1\)) surface structure occurs after the overall annealing cycle.\par

Structural analysis reveals that the annealing cycles lead to homogenous layers of Ni$_x$Fe$_{3-x}$O$_4$. In case of sample A consideration of an additional NiO surface layer on the surface leads to the best agreement between calculated and experimentally observed XRR and SR-XRD results. For sample B SR-XRD indicates a strongly distorted structure with a vertical lattice constant of $c$~=~0.8334\,nm whereas the vertical lattice constant of sample A is close to that of bulk NiFe$_2$O$_4$.\par

These findings are supported by the HAXPES experiments. Firstly, the formation of Fe$^{3+}$ upon annealing is confirmed by the Fe 2p core level HAXPES data. Secondly, for sample B the shape of the Ni 2p-HAXPES indicate the formation of an inverse spinel ferrite, whereas in case of sample A NiO characteristic features first diminish after annealing at 600~$^\circ$C and re-appear after the entire annealing cycle at 800~$^\circ$C. This may be associated with the much thicker initial NiO layer of sample A maybe leading to NiO rich grains at the interface or NiO clusters at the sample surface. Thirdly, we determined a Ni:Fe ratio of 1:2.6 for sample B, the resulting stoichiometry of sample B is Ni$_{0.83}$Fe$_{2.17}$O$_4$. For sample A an increasing amount of Ni$^{2+}$ ions with increasing annealing temperature is found due to the Ni diffusion to the surface.\par

We employed XMCD to study the internal magnetic properties of the thin films resulting from the Ni interdiffusion process. In excellent agreement to complementary charge transfer multiplet simulations we find a strong increase of Fe$^{3+}_{tet}$ coordinated cation fraction (around 50\%) compared to stoichiometric Fe$_3$O$_4$, resulting in very small Fe net magnetic moments as determined from the experimental XMCD data by applying the sum rules. The magnetic properties after the annealing cycle are in both samples dominated by the contribution of the Ni$^{2+}$ ions, which exhibit magnetic moments of 0.56$\mu_B$/f.u. (sample A) and 1.03$\mu_B$/f.u. (sample B). The latter value corresponds quite well to the value very recently reported for a stoichiometric NiFe$_2$O$_4$ thin film.\cite{kle14} The lower value found for sample A can be explained by the formation of (antiferromagnetic) NiO-rich islands or clusters at the surface of the sample which contribute to the Ni L$_{2,3}$-XAS signal but not to the corresponding XMCD. Finally, performed element specific hysteresis loops on sample B find a rather small in-plane coercive field of around 0.02~T. This is a further indication for the formation a quite high quality NiFe$_2$O$_4$-like thin film by means of thermal interdiffusion of Ni$^{2+}$ ions into magnetite from Fe$_3$O$_4$/NiO bilayers.

\begin{acknowledgments}
Financial support by the Deutsche Forschungsgemeinschaft (DFG) (KU2321/2-1 and KU3271/1-1) is gratefully acknowledged. We thank Diamond Light Source for access to beamline I09 (SI10511-1) that contributed to the results presented here. Additionally, parts of this research were carried out at the light source PETRA III at DESY. We would like to thank F. Bertram for assistance using beamline P08. Furthermore, part of this work has been performed at the Advanced Light Source, which is supported by the Director, Office of Science, Office of Basic Energy Sciences, of the U.S. Department of Energy under Contract No. DE-AC02-05CH11231.
\end{acknowledgments}

\bibliography{Diamond}

\end{document}